# Spatial and Magnetic Confinement of Massless Dirac Fermions

Ya-Ning Ren[1,§], Qiang Cheng[2,3,§], Chao Yan[1], Ke Lv[1], Mo-Han Zhang[1], Qing-Feng Sun[3,4,5,*], and Lin He[1,*]

[1]Center for Advanced Quantum Studies, Department of Physics, Beijing Normal University, Beijing, 100875, People's Republic of China
[2]School of Science, Qingdao University of Technology, Qingdao, Shandong 266520, China
[3]International Center for Quantum Materials, School of Physics, Peking University, Beijing, 100871, China
[4]Collaborative Innovation Center of Quantum Matter, Beijing 100871, China
[5]Beijing Academy of Quantum Information Sciences, West Bld. #3, No. 10 Xibeiwang East Road, Haidian District, Beijing 100193, China

[§]These authors contributed equally to this work.
*Correspondence and requests for materials should be addressed to Qing-Feng Sun (e-mail: sunqf@pku.edu.cn) and Lin He(e-mail: helin@bnu.edu.cn).

**The massless Dirac fermions and the ease to introduce spatial and magnetic confinement in graphene provide us unprecedented opportunity to explore confined relativistic matter in this condensed-matter system. Here we report the interplay between the confinement induced by external electric fields and magnetic fields of the massless Dirac fermions in graphene. When the magnetic length $l_B$ is larger than the characteristic length of the confined electric potential $l_V$, the spatial confinement dominates and a relatively small critical magnetic field splits the spatial-confinement-induced atomic-like shell states by switching on a π Berry phase of the quasiparticles. When the $l_B$ becomes smaller than the $l_V$, the transition from spatial confinement to magnetic confinement occurs and the atomic-like shell states condense into Landau levels (LLs) of the Fock-Darwin states in graphene. Our experiment demonstrates that the spatial confinement dramatically changes the energy spacing between the LLs and generates large electron-hole asymmetry of the energy spacing between the LLs. These results shed light on puzzling observations in previous experiments, which hitherto remained unaddressed.**

Graphene, with massless Dirac fermions as the low-energy quasiparticles, provides a unique solid-state system to explore many oddball effects of quantum-relativistic matter [1-5]. More importantly, the one-atom-thick structure of graphene is uniquely amenable to introduce electric and magnetic fields to tune the behaviors of the relativistic fermions. For example, recent experiments demonstrated the confinement of the massless Dirac fermions in graphene into quasibound states via whispering-gallery modes by using circular electronic junctions (graphene quantum dot) [6-11], which enable the demonstration of the so-called Klein tunneling [12] at the atomic scale. By further introducing the magnetic fields, it is interesting to find that a small critical magnetic field could lift the angular momentum quantum numbers of the quasibound states by switching on a $\pi$ Berry phase [13] and a large magnetic field can reshape the electric potentials into wedding cake-like structures through electron-electron interactions [14]. These results indicate that the magnetic fields can strongly modify the effects of the spatial confinements and realize more exotic electronic properties in graphene.

In this Letter, we use scanning tunneling microscopy (STM) to directly probe the interplay between the spatial and magnetic confinements of massless Dirac fermions in graphene. In our experiment, a circular electronic junction confirms the massless Dirac fermions into atomic-like shell states and a relatively small critical magnetic field splits the quasibound states with different angular momentum quantum numbers by switching on a $\pi$ Berry phase. Not only the spatial confinement is modified by the magnetic field, our result indicates that the spatial confinement also can strongly affect the effects of the magnetic confinement. When the magnetic length $l_B$ changes from $l_B > l_V$ to $l_B < l_V$ ($l_V$, the characteristic length of the confined electric potential), the transition from spatial confinement to magnetic confinement occurs and the quasibound states gradually condense into Landau levels (LLs) of the Fock-Darwin states in graphene [15]. Our experiments demonstrate that the spatial confinement can change the energy spacing between the LLs and introduce large electron-hole asymmetry between the energy spacing of the LLs.

Our experiments were carried out on monolayer graphene on a 0.7% Nb-doped SrTiO$_3$(001) substrate by using a high-magnetic-field STM at $T$ = 4.2 K (see Supplemental Material for details [16]). The work function difference between the metal STM tip and graphene leads to an electric field acting on graphene and results in the confining potential on the hole, *i.e.*, a circular electronic *p-n* junction, in graphene below the tip [6,13,14,17-22], as schematically shown in Fig. 1(a). During the STM measurements, the confining potential also depends on the tunneling bias between graphene and the probe tip $V_b$, as demonstrated explicitly in Ref. [6]. This allows us to tune the spatial confining potential by slightly varying the doping of the graphene monolayer (see Supplemental Material Fig. S2 for details [16]). Figure 1(b) shows two representative scanning tunneling spectroscopy (STS) spectra recorded in two different regions of the monolayer graphene (labelled as D1 and D2 respectively). Due to the different charge transfer between the substrate and the graphene, the Dirac points $E_D$ in the D1 and D2 regions are at about (147 ± 5) meV and (74 ± 5) meV, respectively. Although the work function difference between the metal STM tip and graphene can generate band bending of graphene beneath the tip in both the regions, the dependence of the band bending on the $V_b$ makes us almost unable to detect the effect of the confining potential in the zero-magnetic-field STS measurements in the D2 region (see Supplemental Material for details [16]). Therefore, no resonance can be measured in tunneling spectrum of the D2 region [bottom panel of Fig. 1(b)]. Whereas, clear whispering-gallery resonances, corresponding to tunneling into the circular *p-n* junction modes at energy $\varepsilon = \mu_0 + eV_b$ ($\mu_0$ is the local Fermi level) [6], are observed in the tunneling spectrum of the D1 region [top panel of Fig. 1(b)]. The energy spacing between the quasibound states is consistent well with that observed previously in the tip-induced graphene resonators [6]. The ability to tune the spatial confining potential allows us to systematically explore the interplay between the spatial and magnetic confinements of the massless Dirac fermions in monolayer graphene, as illustrated below.

A prominent effect of the magnetic fields on the quasibound states of the massless Dirac fermions is the π shift of the Berry phase of the confined quasiparticles [13,23].

At zero magnetic field, the quasibound states with opposite angular momenta ±m are degenerate [Fig. 1(c), top] and their corresponding momentum-space loops [Fig. 1(d), blue curve] do not enclose the origin. Therefore, the Berry phase of the quasibound states is zero. A small magnetic field breaks the time-reversal symmetry and bends the paths of the +m and −m quasibound states in opposite directions (in our experiment m=1/2, because only the angular momenta ±1/2 states have nonzero wave function amplitude in the center of the resonators). At a critical magnetic field $B_c$, the Lorentz force can twist the orbit with angular momentum antiparallel to the magnetic field into a skipping orbit with loops [Fig. 1(c), bottom]. Then, the momentum-space loop encircles the Dirac point [Fig. 1(d), red curve] and the Berry phase becomes π (the Berry phase of the orbit with angular momentum parallel to the magnetic field is still zero). The π shift of the Berry phase will result in a sudden and large decrease in the energy of the corresponding angular-momentum (+m) quasibound states [13,23]. Such a feature is clearly observed in our STS measurements in the D1 region, as shown in Fig. 1(e). A critical magnetic field lift the angular momenta ±1/2 degeneracy of the quasibound states and, then, the spacing between the new states is about one-half the spacing at zero magnetic field [Fig. 1(e), left].

To further understand our experimental result, we theoretically calculated local density of states (LDOS) in the center of the graphene resonator as a function of magnetic fields and energy. In our calculations, the following Hamiltonian is adopted

$$H = v_F \boldsymbol{\sigma} \cdot \boldsymbol{\pi} + U(r) + E_D. \qquad (1)$$

Here $v_F$ is the Fermi velocity, $E_D$ is the energy of the Dirac point, $\boldsymbol{\sigma} = (\sigma_x, \sigma_y)$ are pseudospin Pauli matrices and $\boldsymbol{\pi} = (\pi_x, \pi_y)$ are the kinematic momenta with $\pi_{x,y} = -i\hbar \partial_{x,y} - eA_{x,y}$. The symbol $\hbar$ is Plank's constant divided by $2\pi$. The axial gauges, $A_x = -By/2$ and $A_y = Bx/2$ with magnetic field $B$, are used to preserve the rotational symmetry. For simplicity, we consider a fixed circular-shaped p-n junction with a parabolic confining potential $U(r) = -\kappa r^2$. Due to the rotational symmetry, the eigenstates of $H$ can be expressed as $\psi_m(r,\theta) = \frac{e^{im\theta}}{\sqrt{r}}(u_1(r)e^{-i\theta/2}, iu_2(r)e^{i\theta/2})^T$, with the angular momentum quantum number $m$ a half-integer number. The LDOS of the graphene resonator can be expressed as the sum of $m$-state contributions [6,23]. The $\pi$ shift of Berry phase and the sudden decrease of energy, as discussed above, can be explained using the Einstein-Brillouin-Keller rule [24-26] starting from the Hamiltonian here (see Supplemental Material for details [16]).

Figure 1(e) shows the simulated LDOS for the D1 region with $v_F = 1.24 \times 10^6$ m/s, $\kappa = 30$ eV/$\mu$m$^2$ and $E_D = 146.5$ meV. Obviously, the results, including the energy spacing $\Delta E$ of the quasibound states at zero magnetic field, the sudden and large reduce in the energy of the angular momentum +1/2 states above the critical magnetic field, and the value of the critical magnetic field $B_C = 2\hbar m\kappa/[e(E_D - E)] \approx 0.3$T, agree quite well with our experimental results. The good agreement between the experiment and theory indicates that the tip-induced potential is weakly dependent on the tunneling bias in the measured range, -25 mV $< V_b <$ 100 mV. According to both the experimental results and theoretical calculation, a characteristic length scale for the confining potential $l_V = \hbar v_F/\Delta E$ can be estimated as about 20 nm. Applying a magnetic field $B$ will confine holes in a region of the magnetic length $l_B = [\hbar/(eB)]^{1/2}$. Therefore, for the case that the $l_V$ is smaller than the $l_B$, the spatial confinement dominates and we can observe the quasibound states, as shown in Fig. 1(e). When the $l_B$ becomes smaller than the $l_V$, a transition from spatial confinement to magnetic confinement occurs and the LLs of massless Dirac fermions begin to be observed. In the D1 region, the characteristic Landau quantization of monolayer graphene is observed for $B > 1.4$ T, which also agrees well with the estimated $l_V$ above. In the D2 region, the effect of the tip-induced confining potential in the zero-field measurement is almost negligible. Therefore, we observe the characteristic LLs of monolayer graphene for the magnetic field on the order of 0.1 T [Fig. 1(f), left]. Such a feature is also reproduced quite well in our calculation of the Landau quantization in monolayer graphene when we consider a very weak confining potential with $\kappa = 9$ eV/$\mu$m$^2$ and $E_D = 74$ meV, as well as $v_F = 1.1 \times 10^6$ m/s. [Fig. 1(f), right, see Supplemental Material [16] for details].

To further explore the effect of spatial confinement on the magnetic confinement of the massless Dirac fermions, the STS spectra in the D1 and D2 regions in a large range of magnetic fields and sample bias $V_b$ are measured, as shown in left panels of Figs. 2(a) and 2(b) respectively. The theoretical Landau quantization of monolayer graphene with considering the tip-induced potentials, as used in the Figs. 1(e) and 1(f), is also calculated for comparison [right panels of Figs. 2(a) and 2(b)]. Even though the tip-induced potential in the D2 region cannot generate resonances in zero-magnetic-field spectra, the spatial confinement strongly affects the magnetic confinement in several aspects in both the D1 and D2 regions. First, the tip-induced potential $-\kappa r^2$ is a repulsive potential for the electron, which inhibits the formation of quasibound states for the electron in the range $B^2 < \frac{16\kappa}{(v_F e)^2}(eV_b - E_D)$ [see Supplemental Material [16] for details]. Therefore, the LLs can appear at the electron band when $B^2 > \frac{16\kappa}{(v_F e)^2}(eV_b - E_D)$, *i.e.*, the magnetic confinement overcomes the repulsive potential, as shown in Fig. 2(a) and 2(b). In pristine graphene monolayer, we can observe well-defined Landau quantization of the massless Dirac fermions, $E_n = \text{sgn}(n)\sqrt{2e\hbar v_F^2 |n| B} + E_0$, with the LL number $n$ = 0, ±1, ±2, …($n > 0$ corresponds to electrons and $n < 0$ to holes), in the presence of high magnetic field. The energy of the zero LL should be independent of the magnetic fields in the absence of the spatial confinement (see Supplemental Material Fig. S3 for details [16]). The second observable effect of the spatial confinement on the Landau quantization is that the zero LL varies slightly with the magnetic field, as shown in Fig. 2(a) and 2(b). The variation of the zero LL in the D1 region is much more pronounced than that in the D2 region due to the much stronger confining potential. Our theoretical results, with considering both the spatial confinement and magnetic confinement, reproduce quite well the main features, including the Landau quantization and the slight variation of the zero LL, of our experiment. In our experiment, we observed slight splitting of the LLs, which may arise from electron-electron interactions that are not taken into account in our theoretical calculations. Besides the LLs, we observe several lines intersecting the LLs at the Fermi level and progressing upward at sharp angles, as shown in Fig. 2. These charging lines arise from Coulomb charging effects, which are generated by shifting quasiparticles in the region beneath the tip into the gaps between the LLs of the adjacent regions [14,17-20].

In pristine monolayer graphene, the slope of the data $E_n$ *versus* $\text{sgn}(n)\sqrt{|n|B}$ directly reflects the Fermi velocity of the massless Dirac fermions and should be a constant. Because of the existence of the tip-induced spatial potential, there are pronounced deviations between the observed Landau quantization in both the D1 and D2 regions and that expected in pristine monolayer graphene, as plotted in Figs. 3(a) and 3(c). To qualitatively show the deviation, the Fermi velocity at each magnetic field is extracted according to the energy spacing of the LLs, $v_F = (E_n - E_0)/[\text{sgn}(n)\sqrt{2e\hbar|n|B}]$, and is plotted in Figs. 3(b) and 3(d) for the D1 and D2 regions, respectively. The spacing of the LLs for holes is systematically larger than that for electrons, *i.e.*, the deduced "Fermi velocity" for holes is larger than that for electrons, because that the tip-induced potential is a confining potential for the hole but it is a repulsive potential for the electron. In the D2 region, the electron-hole asymmetry, as defined by $v_F^h/v_F^e$, could be as large as 110%. In the D1 region, the electron-hole asymmetry increases to about 130% due to the relatively stronger tip-induced potential. Generally, the electron-hole asymmetry is much more pronounced in small magnetic fields. Such a result is quite reasonable because that the spatial confinement is expected to change the spacing of the LLs, especially for the case that the $l_V$ is comparable to the $l_B$. In previous STM studies of the Landau quantization in monolayer graphene, similar large electron-hole asymmetry has been frequently observed and is mainly attributed to the enhanced next-nearest-neighbor hopping [19,20,24-27]. To account for the electron-hole asymmetry observed in the experiment, an unreasonable large next-nearest-neighbor hopping ~ 1 eV is needed to be used as a fitting parameter in the tight-binding model.

By taking into account the effects of the tip-induced potential, our calculation can reproduce quite well the above experimental phenomena (Figs. 2 and 3), indicating that the large electron-hole asymmetry observed in STM studies is mainly arising from the tip-induced potential. In our experiment, the tip-induced potential promotes and resists the formation of the Landau quantization of the hole and electron respectively, resulting in the giant electron-hole asymmetry.

To further explore the effects of spatial confinement on the magnetic confinement of the massless Dirac fermions, we calculated the Landau quantization of monolayer graphene with different strengths of the confining potential (see Supplemental Material [16] for details). Figure 4 summarizes the calculated Fermi velocity for electrons and holes as a function of the confining potential. The *n-p-n* and *p-n-p* junctions have opposite confining effects for the electrons and holes, therefore, generating opposite electron-hole asymmetry on the energy spacing, *i.e.*, the Fermi velocity, of the LLs. At a fixed magnetic field, the electron-hole asymmetry increases with increasing the strength of the confining potential, which is consistent with our experimental results obtained in the D1 and D2 regions. For a fixed confining potential, the effect of spatial confinement on the Landau quantization becomes more obvious with decreasing the magnetic fields. Such a result is quite reasonable because the characteristic length of the confining potential $l_V$ is comparable to the magnetic length $l_B$ in small magnetic fields and also agrees well with our experimental results.

In summary, the interplay between the spatial and magnetic confinements of massless Dirac fermions in graphene is systematically explored in this work. Our result demonstrates that a relatively small critical magnetic field lifts the angular momentum degeneracy of the spatial-confined quasibound states by switching on a $\pi$ Berry phase. In turn, the spatial potential also strongly modifies the magnetic-field-induced Landau quantization and generates giant electron-hole asymmetry in the energy spacing between the LLs.

**Acknowledgments**

This work was supported by the National Natural Science Foundation of China (Grant Nos. 11974050, 11674029, 11921005) and National Key R and D Program of China (Grant No. 2017YFA0303301). L.H. also acknowledges support from the National Program for Support of Top-notch Young Professionals, support from "the Fundamental Research Funds for the Central Universities", and support from "Chang Jiang Scholars Program".

**Reference:**

[1] A. K. Geim, K. S. Novoselov, *Nature Mater.* **6**, 183-191 (2007).

[2] A. H. Castro Neto, N. M. R. Peres, K. S. Novoselov, A. K. Geim, *Rev. Mod. Phys.* **81**, 109-162 (2009).

[3] M. A. H. Vozmediano, M. I. Katsnelson, F. Guinea, *Physics Rep.* **496**, 109-148 (2010).

[4] S. Das Sarma, S. Adam, E. Hwang, E. Rossi, *Rev. Mod. Phys.* **83**, 407-470 (2011).

[5] M. Goerbig, *Rev. Mod. Phys.* **83**, 1193-1243 (2011).

[6] Y. Zhao, J. Wyrick, F. D. Natterer, J. F. Rodriguez-Nieva, C. Lewandowski, K. Watanabe, T. Taniguchi, L. S. Levitov, N. B. Zhitenev, and J. A. Stroscio, *Science* **348**, 672–675 (2015).

[7] C. Gutiérrez, L. Brown, C.-J. Kim, J. Park, and A. N. Pasupathy, *Nat. Phys.* **12**, 1069–1075 (2016).

[8] J. Lee, D. Wong, J. Velasco Jr, J. F. Rodriguez-Nieva, S. Kahn, H.-Z. Tsai, T. Taniguchi, K. Watanabe, A. Zettl, F. Wang, L. S. Levitov, and M. F. Crommie, *Nat. Phys.* **12**, 1032–1036 (2016).

[9] K.-K. Bai, J.-J. Zhou, Y.-C. Wei, J.-B. Qiao, Y.-W. Liu, H.-W. Liu, H. Jiang, and L. He, *Phys. Rev. B* **97**, 045413 (2018).

[10] Z.-Q. Fu, K. K. Bai, Y.-N. Ren, J.-J. Zhou, and L. He, *Phys. Rev. B* **101**, 235310 (2020).

[11] Z.-Q. Fu, Y.-T. Pan, J.-J. Zhou, K.-K. Bai, D.-L. Ma, Y. Zhang, J.-B. Qiao, H. Jiang, H. Liu, and L. He, *Nano Lett.* **20**, 6738 (2020).

[12] M. I. Katsnelson, K. S. Novoselov, A. K. Geim, *Nat. Phys.* **2**, 620-625 (2006).

[13] F. Ghahari, D. Walkup, C. Gutiérrez, J. F. Rodriguez-Nieva, Y. Zhao, J. Wyrick, F. D. Natterer, W. G. Cullen, K. Watanabe, T. Taniguchi, L. S. Levitov, N. B. Zhitenev, and J. A. Stroscio, *Science* **356**, 845 (2017).

[14] C. Gutiérrez, D. Walkup, F. Ghahari, C. Lewandowski, J. F. Rodriguez-Nieva, K. Watanabe, T. Taniguchi, L. S. Levitov, N. B. Zhitenev, and J. A. Stroscio, *Science* **361**, 789–794 (2018).

[15] H.-Y. Chen, V. Apalkov, T. Chakraborty, *Phys. Rev. Lett.* **98**, 186803 (2007).

[16] See the Supplemental Material for sample preparation, more experimental data, details of calculation and discussion.

[17] N. M. Freitag, L. A. Chizhova, P. Nemes-Incze, C. R. Woods, R. V. Gorbachev, Y. Cao, A. K. Geim, K. S. Novoselov, J. Burgdörfer, F. Libisch, and M. Morgenstern, *Nano Lett.* **16**, 5798 (2016).

[18] N. M. Freitag, T. Reisch, L. A. Chizhova, P. Nemes-Incze, C. Holl, C. R. Woods, R. V. Gorbachev, Y. Cao, A. K. Geim, K. S. Novoselov, J. Burgdörfer, F. Libisch, and M. Morgenstern, *Nat. Nanotechnol.* **13**, 392 (2018).

[19] S.-Y. Li, M.-X. Chen, Y.-N. Ren, H. Jiang, Y.-W. Liu, and L. He, *2D Mater.* **6**, 031005 (2019).

[20] S.-Y. Li, Y. Su, Y.-N. Ren, and L. He, *Phys. Rev. Lett.* **124**, 106802 (2020).

[21] J. Velasco Jr, J. Lee, D. Wong, S. Kahn, H.-Z. Tsai, J. Costello, T. Umeda, T. Taniguchi, K. Watanabe, A. Zettl, F. Wang, M. F. Crommie, *Nano Lett.* **18**, 5104 (2018).

[22] Y.-W. Liu, Z. Hou, S.-Y. Li, Q.-F. Sun, L. He, *Phys. Rev. Lett.* **124**, 166801 (2020).

[23] J. F. Rodriguez-Nieva and L. S. Levitov, *Phys. Rev. B* **94**, 235406 (2016).

[24] Z. Hou, Y.-F. Zhou, X. C. Xie, and Q.-F. Sun, *Phys. Rev. B* **99**, 125422 (2019).

[25] A. Einstein, *Deutshe Phys. Ges.* **19**, 82 (1917).

[26] A. D. Stone, *Phys. Today* **58**, 37 (2005).

[27] A. Luican, G. Li, A. Reina, J. Kong, R. R. Nair, K. S. Novoselov, A. K. Geim, E. Y. Andrei, *Phys. Rev. Lett.* **106**, 126802 (2011).

[28] K.-K. Bai, Y.-C. Wei, J. B. Qiao, S.-Y. Li, L.-J. Yin, W. Yan, J.-C. Nie, and L. He, *Phys. Rev. B* **92**, 121405(R) (2015).

[29] Y. Zhang, S.-Y. Li, H. Huang, W. T. Li, J. B. Qiao, W.-X. Wang, L.-J. Yin, W. H. Duan, and L. He, *Phys. Rev. Lett.* **117**, 166801 (2016).

[30] S.-Y. Li, K.-K. Bai, W. J. Zuo, Y.-W. Liu, Z.-Q. Fu, W.-X. Wang, Y. Zhang, L.-J. Yin, J.-B. Qiao, and L. He, *Phys. Rev. Appl.* **9**, 054031 (2018).

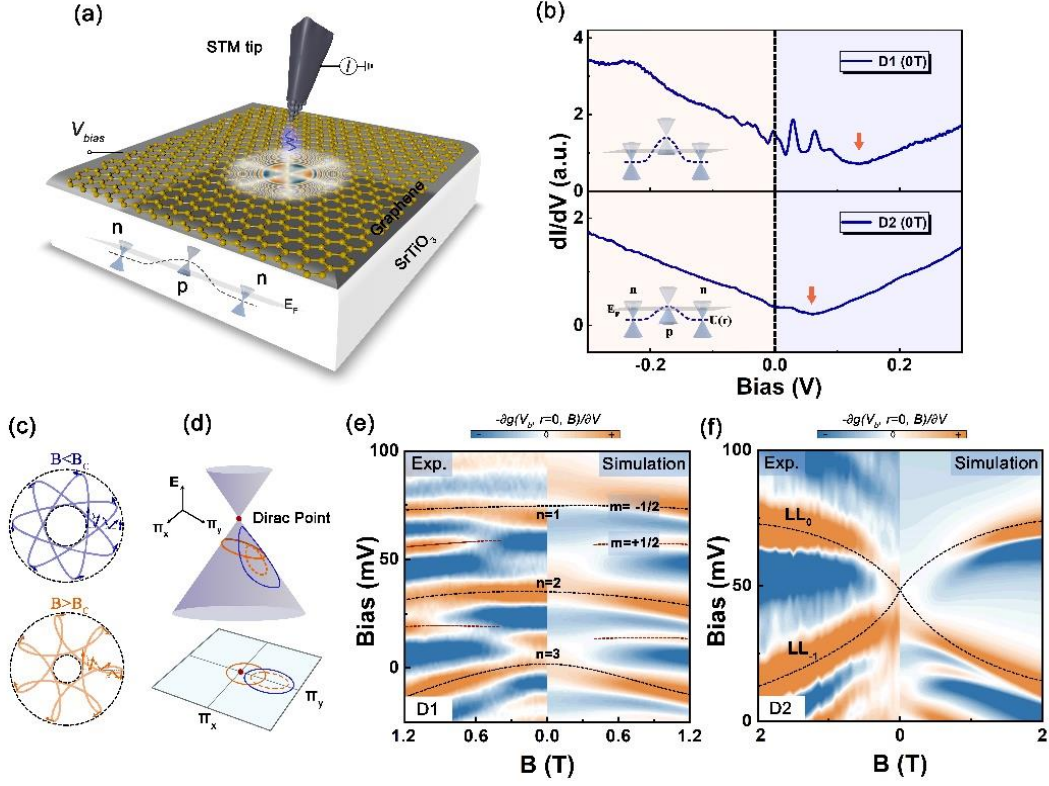

**Fig. 1.** STS measurements of the tip-induced circular graphene resonators. (a) Sketch of the STM setup. The electric field of the STM tip induces band bending, generating confining potential to confine the massless Dirac fermions via whispering-gallery modes. (b) STS spectra of monolayer graphene with different doping. Dirac points are marked with orange arrows. Insets are the schematic of tip-induced circular n–p–n junctions. (c) Schematic diagrams of charge orbits in the circular graphene resonators under different applied magnetic fields, corresponding to $B<B_C$ (top panel) and $B>B_C$ (Bottom panel). (d) Schematic charge trajectories in momentum space on the Dirac cone for magnetic fields below (blue) and above (orange) the critical magnetic field $B_C$. The orange solid and dashed lines represent $m = +1/2$ and $m = -1/2$ modes, respectively. (e) and (f) Differential conductance maps versus magnetic fields $B$. Panel (e) shows the sudden jumping in energy of the quasibound states $n = 1$ and $n = 2$ modes. When magnetic field is larger than the critical magnetic field $B_C$, there is a large and sharp jump in energy of the $m = +1/2$ state. Panel (f) shows spectra of the D2 region with a very weak spatial confinement. No resonances can be observed and the Landau levels are clearly observed in magnetic fields on the order of 0.1 T.

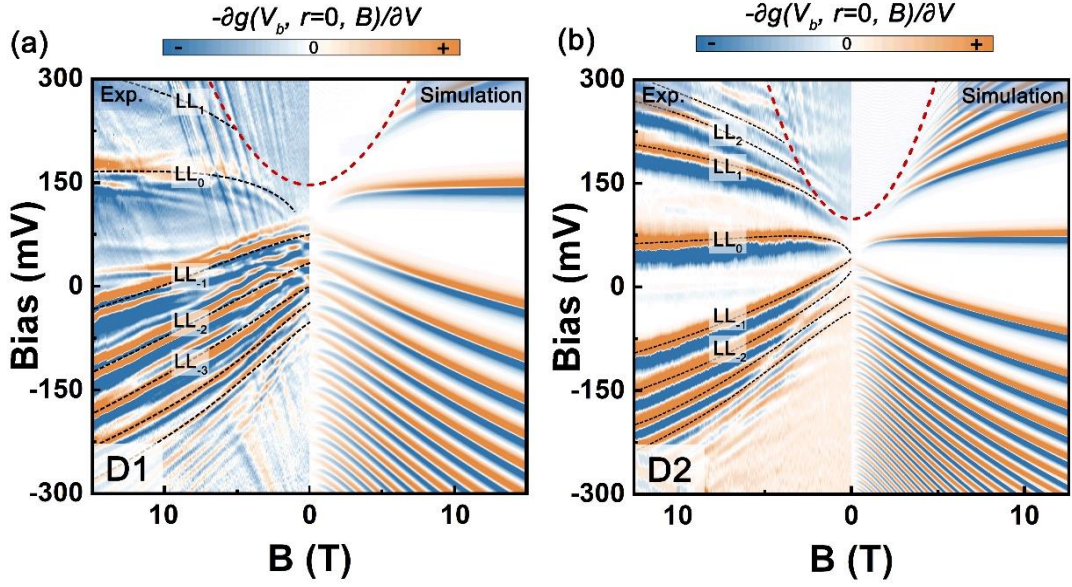

**Fig. 2.** Differential conductance maps versus magnetic field B in the two regions with different spatial confinements. Panels (a) (left) and (b) (left) correspond to the spectra recorded in the D1 and D2 regions respectively. Landau levels of massless Dirac fermions are clearly observed in the spectra recorded in high magnetic fields. Panels (a) (right) and (b) (right) show the calculated Landau quantization of monolayer graphene in the presence of different spatial confinements. The red dotted line is $B^2 = \frac{16\kappa}{(v_F e)^2}(eV_b - E_D)$. The parameters for the theoretical simulation are taken as $v_F = 1.24 \times 10^6 m/s$, $\kappa = 30 eV/\mu m^2$, $E_D = 146.5 meV$ for (a) and $v_F = 1.1 \times 10^6 m/s$, $\kappa = 9 eV/\mu m^2$, $E_D = 74 meV$ for (b).

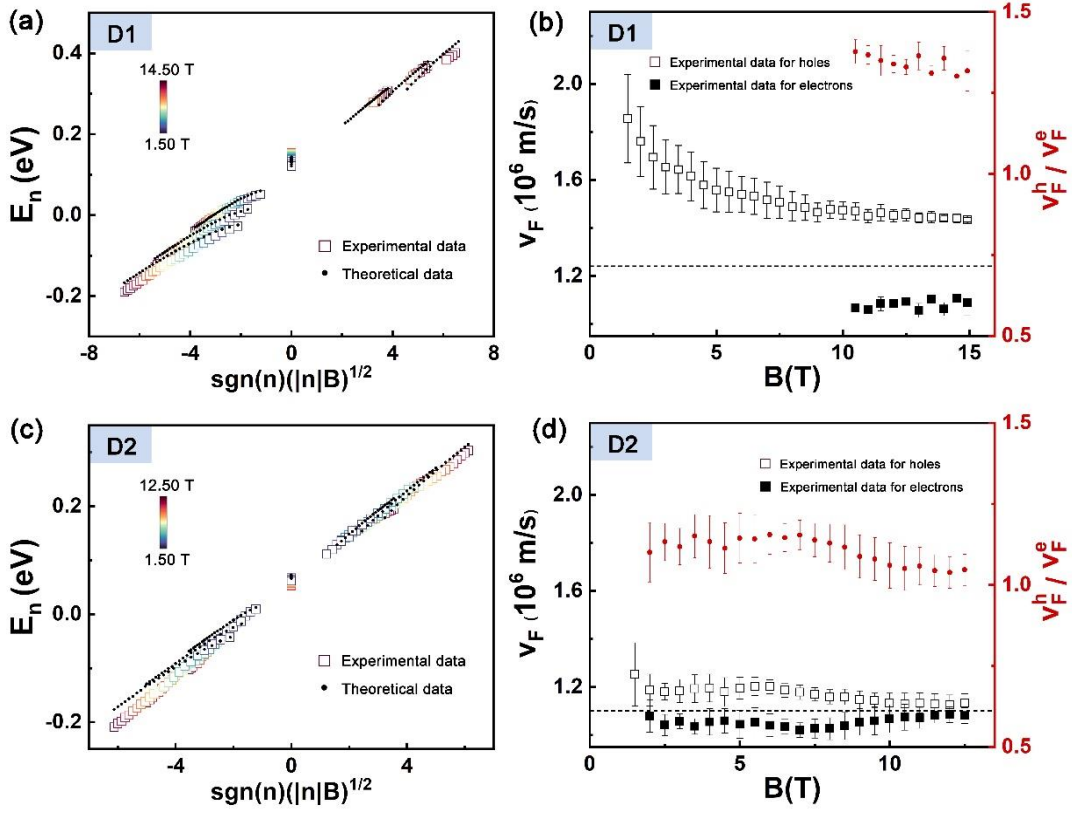

**Fig. 3.** Spatial confinements induced Fermi velocity renormalization and electron–hole asymmetry. (a) and (c) The Landau level energies against $sgn(n)(|n|B)^{1/2}$ recorded in the D1 and D2 regions respectively. The squares represent the experimental data, which are recorded with an interval of 0.5T in magnetic fields. Black dots correspond to theoretical results extracted from Fig. 2. (b) and (d) Fermi velocities and electron-hole asymmetry of the Fermi velocity ($v_F^h/v_F^e$) as a function of magnetic fields. The theoretical result without considering the confining potential is also plotted (dashed lines) for comparison. There is no electron-hole asymmetry and the Fermi velocity is independent of the magnetic field when there is no confining potential.

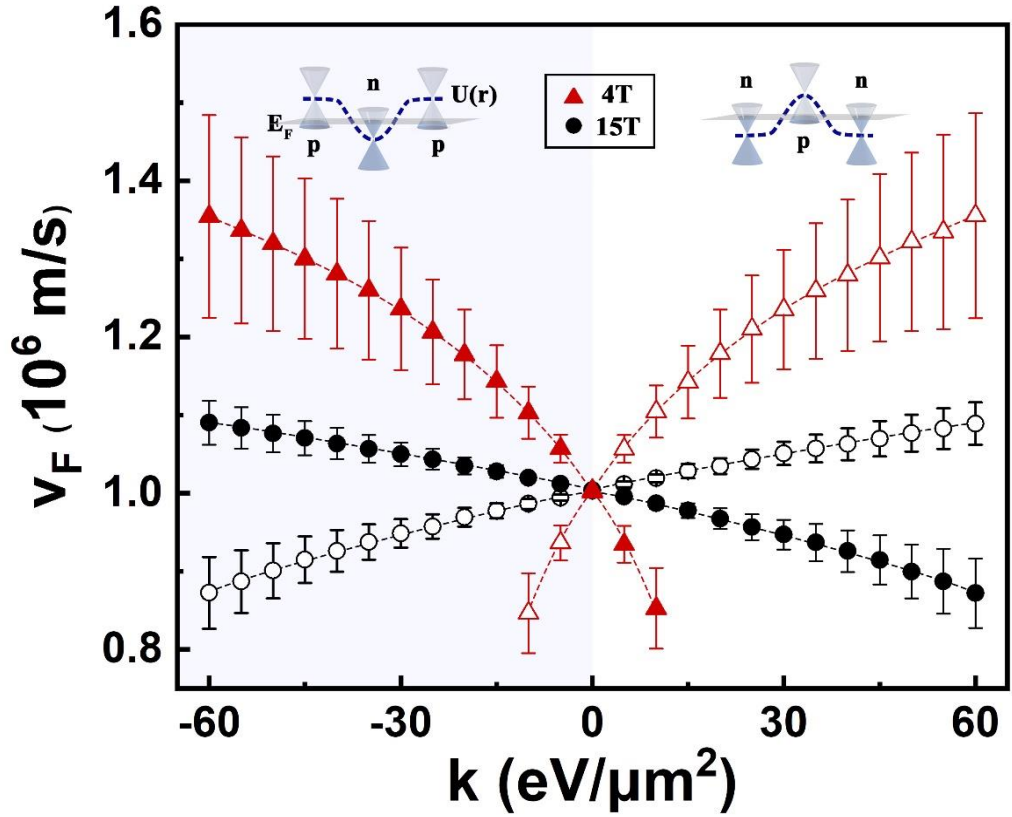

**Fig. 4.** The theoretical Fermi velocity for the electrons and holes as a function of the confining potential. The insets schematically show the circular confining potentials. The *n-p-n* and *p-n-p* confining junctions have opposite effects on the Landau quantization for electrons (solid symbols) and holes (open symbols), therefore, generating opposite electron-hole asymmetry.